\documentclass{nature}

\usepackage{amsmath}
\usepackage{amsfonts}
\usepackage{amssymb}
\usepackage{graphicx}
\usepackage{epstopdf}
\newcommand{\bra}[1]{\left\langle#1\right|}
\newcommand{\ket}[1]{\left|#1\right\rangle}
\newcommand{\braket}[2]{\langle #1 | #2 \rangle}
\newcommand{\hlf}{\frac{1}{2}}
\newcommand{\us}{\uparrow}
\newcommand{\ds}{\downarrow}

\title{Quantum spin transistor with a Heisenberg spin chain}

\author{O.~V. Marchukov$^{1}$, A.~G. Volosniev$^{1,2}$, M. Valiente${}^{3}$,  
D. Petrosyan,${}^{4}$ \& N.~T. Zinner${}^{1}$}

\begin{document}

\maketitle

\begin{affiliations}
 \item Department of Physics and Astronomy, Aarhus University, DK-8000 Aarhus C, Denmark
 \item Institut f{\"u}r Kernphysik, Technische Universit{\"a}t Darmstadt, 64289 Darmstadt, Germany
 \item SUPA, Institute of Photonics and Quantum Sciences, Heriot-Watt University, Edinburgh EH14 4AS, United Kingdom
 \item Institute of Electronic Structure and Laser, FORTH, GR-71110 Heraklion, Crete, Greece
\end{affiliations}

\begin{abstract}
We propose and analyze a scheme for conditional state transfer in 
a Heisenberg $XXZ$ spin chain which realizes a quantum spin transistor. 
In our scheme, the absence or presence of a control spin excitation 
in the central gate part of the spin chain results 
in either perfect transfer of an arbitrary state of a target spin 
between the weakly coupled input and output ports, or its complete 
blockade at the input port. 
We also present a possible realization of the corresponding spin chain
with a one-dimensional ensemble of cold atoms with strong contact interactions.
\end{abstract}

Starting with the original proposal by Datta and Das of 
a spin-based field-effect transistor \cite{datta1990}, the field of spintronics
\cite{zutic2004}
has explored how the spin degrees of freedom can be used for 
information transfer. More than two decades later this 
research has reached the quantum regime \cite{awschalom2013}. 
One motivation for this is the desire for miniaturization 
which led to the realization of single electron transistors \cite{devoret1998}, or more
generally single-dopant devices \cite{koenraad2011}.
A second motivation is the potential applications in 
quantum information and computation. 
Two guiding proposals 
in the field involve implementation of quantum gate operations
in quantum dots \cite{loss1998} and in doped silicon \cite{kane1998}. 
Shortly thereafter, molecular magnets have also been proposed \cite{leuenberger2001}.
It was subsequently shown that universal quantum computation can 
be realized with just the Heisenberg exchange interaction known from 
quantum magnetism \cite{divincenzo2000,levy2002}.

Here we put forward an novel scheme for a quantum spin transistor that 
may serve as an integral component of quantum information devices. 
Similarly to the quantum computation proposals, it can be implemented
on architectures that realize a Heisenberg spin chain. 
Various physical realizations of spin chains are being actively 
explored for short-range quantum state transfer required to 
integrate and scale-up quantum registers involving many qubits  
\cite{bose2003,bose2008,nikolopoulos2004,christandl2004,petrosyan2010}.
In fact, spin chains of the Heisenberg type have been realized in 
organic and molecular magnets \cite{blundell2004}, quantum 
dots \cite{zwanenburg2013}, various compounds \cite{blundell2004,mourigal2013,sahling2015}, 
Josephson junction arrays \cite{fazio2001}, trapped ions \cite{porras2004,blatt2012},
in atomic chains on surfaces \cite{gambardella2002,hirji2006,khaje2011,khaje2012}, and 
in thin films or narrow magnetic strips that carry spin waves 
\cite{Chumak2014,Chumak2015}.  
Combined with conditional dynamics to realize quantum logic gates, 
spin chains can greatly facilitate large-scale quantum information processing. 

While many different realizations of the coherent spin transistor
may be possible, 
here we focus one such realization in a small 
ensemble of strongly-interacting cold atoms trapped in a tight
one-dimensional potential of appropriate shape. Cold atoms have 
already been used to realize spin chains and observations of 
Heisenberg exchange dynamics \cite{trotzky2008}, spin impurity
dynamics \cite{fukuhara2013a} and magnon bound states \cite{fukuhara2013b} 
have been reported.

Our quantum spin transistor works with  
an arbitrary spin state $\ket{\psi} = \alpha \ket{\ds} +\beta \ket{\us}$ 
at the input port (target spin) which is coherently transferred to the output 
port, if there are no excited spins in the gate, $\ket{0}_{\mathrm{gate}}$.
However, if the gate contains an excited stationary spin (control spin), 
$\ket{1}_{\mathrm{gate}}$, it blocks completely the transfer of the target 
spin state between the input and output ports. 
In other words, we have coherent dynamics for the initial state of 
the system $\ket{\psi}_{\mathrm{in}} \ket{0}_{\mathrm{gate}} \ket{\ds}_{\mathrm{out}} 
\to \ket{\ds}_{\mathrm{in}} \ket{0}_{\mathrm{gate}} \ket{\psi}_{\mathrm{out}}$, 
but complete absence of dynamics for the state
$\ket{\psi}_{\mathrm{in}} \ket{1}_{\mathrm{gate}} \ket{\ds}_{\mathrm{out}}$ 
when the gate contains a single spin excitation. Our scheme thus 
realizes a quantum logic operation and it can be used to obtain spatially 
entangled states of target and control spins, as well as to create 
Schr\"odinger cat states for a large number of target spins.

\section*{Coherent spin transistor}
Consider a chain of $N$ spin-$\hlf$ particles described by the 
Heisenberg $XXZ$ model Hamiltonian ($\hbar =1$)
\begin{equation}
\label{eg:ham}
H = \sum_{j = 1}^{N} h_j \sigma_z^j 
- \frac{1}{2} \sum_{j=1}^{N-1} J_j 
[ \sigma_x^j \sigma_x^{j+1} + \sigma_y^j \sigma_y^{j+1} 
+  \Delta \sigma_z^j \sigma_z^{j+1} ] , 
\end{equation}
where $\sigma_{x,y,z}^{j}$ are the Pauli matrices acting on the $j$th spin, 
$h_j$ determine the energy shifts of the spin-up and spin-down states
playing the role of the local magnetic field, $J_j$ are the nearest-neighbor
spin-spin interactions, and $\Delta$ is the asymmetry parameter:
$\Delta = 0$ corresponds to the purely spin-exchange $XX$ model, 
$\Delta = 1$ to the homogeneous spin-spin interaction $XXX$ model,
while the limit of $\Delta \gg 1$ leads to the Ising model. 
We assume a spatially symmetric spin chain with $J_j = J_{N-j}$
and $h_j = h_{N+1-j}$ with $h_{1,N} = 0$.

\begin{figure}
\centering
\includegraphics[scale=0.7]{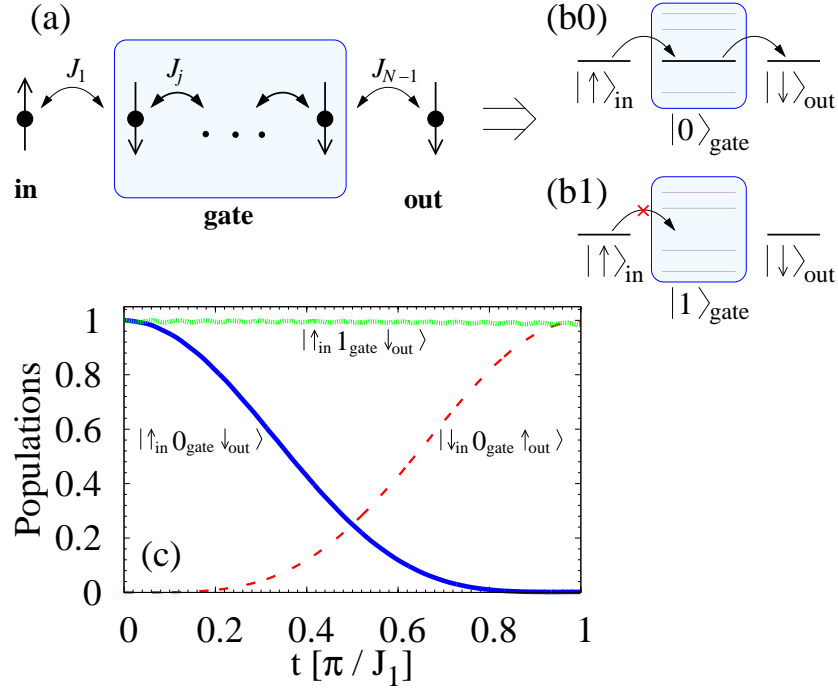}
 \caption{Quantum spin transistor implementation in a spin chain (a) 
where the ``in'' and ``out'' ports are coupled with $J_{1,N-1} \ll J_j$
to the central ``gate'' region. 
The energy levels of the gate, split by $\sim J_j$, are additionally tuned 
by $h_j$, so as to realize resonant transfer of spin excitation between 
the input and output ports when the gate is empty, $\ket{0}_{\mathrm{gate}}$, (b0)
or to block the spin excitation transfer when the gate contains 
an excited (control) spin, $\ket{1}_{\mathrm{gate}}$, (b1).
Dynamics of an $N=4$ spin system with an open or closed gate (c). The 
parameters are $\Delta = -1$, $h=h_-=0$, and $J_2 / J_1 = 17.484$, 
as obtained
for a system of four atoms in the potential of Fig.~\ref{fig:1Dpot}.}
\label{fig:spchTr}
\end{figure}

The input and output ports for the target spin are represented by the 
first $j=1$ and last $j=N$ sites of the chain, see Fig.~\ref{fig:spchTr}(a).
The inner sites $j =2, \ldots, N-1$ constitute the gate which 
may be open or closed for the target spin transfer 
depending on the absence, $\ket{0}_{\mathrm{gate}}$, or presence, 
$\ket{1}_{\mathrm{gate}}$, of a singe control spin excitation.
Ostensibly, the shortest possible spin chain to accommodate a gate 
between the input and output ports would consist of $N=3$ spins.
As we show in the supplementary material, however, the three-spin chain cannot
implement a reusable spin transistor even in the Ising limit 
$|\Delta| \gg J_j$ since the control spin excitation at the 
gate site $j=2$ is not protected from leakage. 
We will therefore illustrate the scheme using a chain of $N=4$ spins, 
with the gate consisting of spins $j=2,3$ coupled to each other via 
the strong exchange constant $J_2 \gg J_{1,3}$. 

Consider a system in the initial state $\ket{\us \ds \ds \ds} \equiv
\ket{\us}_{\mathrm{in}} \ket{0}_{\mathrm{gate}} \ket{\ds}_{\mathrm{out}}$, which 
we aim to efficiently transfer to the final state $\ket{\ds \ds \ds \us} 
\equiv \ket{\ds}_{\mathrm{in}} \ket{0}_{\mathrm{gate}} \ket{\us}_{\mathrm{out}}$ 
using an intermediate resonant state, see Fig.~\ref{fig:spchTr}(b0) and the supplementary material.
The initial and final states have the same energy~\footnote{Throughout this manuscipt, the expectation value of the Hamiltonian in a given state is called the energy of this state. This is the standard convention in quantum optics.} of 
$\tilde{\lambda}_{\us \ds \ds \ds (\ds \ds \ds \us)} = -\hlf J_2 \Delta - 2h$, where $h \equiv h_{2,3}$. 
In turn, the eigenstates of the Hamiltonian~(\ref{eg:ham}) in 
the single-excitation space of strongly-coupled sites $j=2,3$ are 
given by $\ket{G_{\pm}} = \frac{1}{\sqrt{2}} (\ket{\us \ds} \pm \ket{\us \ds})$,
with the corresponding energies $\lambda_{\pm} = \frac{1}{2}J_2\Delta \mp J_2$
split by $2J_2$. Then, by a proper choice of the magnetic field, 
$h=h_{\pm} \equiv \pm \hlf J_2 (1 \mp \Delta)$, we can tune the energy 
of one of the intermediate states $\ket{\ds G_{\pm} \ds}$ into resonance 
with the initial $\ket{\us \ds \ds \ds}$ and final $\ket{\ds \ds \ds \us}$
states (e.g., for $h_{-}$ the resonant state is $\ket{\ds G_{-} \ds}$).
Simultaneously, the other intermediate state does not participate 
in the transfer since its energy is detuned by $2J_2$ which is much larger 
than the coupling rate $J_1/\sqrt{2}$ of the initial and final states 
to the intermediate states. The transfer time of the spin excitation 
between the initial and final states via a single resonant intermediate 
state is $t_{\mathrm{out}} = \pi/J_1$ [Fig.~\ref{fig:spchTr}(c)].  

Next, we place a single spin excitation in one of the eigenstates 
$\ket{G_{\pm}}$ of the gate. To be specific, for
the magnetic field $h=h_-$ we place the control
spin in the state $\ket{G_{+}} = \ket{1}_{\mathrm{gate}}$. Then the control spin 
cannot leak out of the gate region and therefore it is stationary.
Moreover, if we place a target spin-up at the input port, 
the resulting state $\ket{\us G_{+} \ds} \equiv 
\ket{\us}_{\mathrm{in}} \ket{1}_{\mathrm{gate}} \ket{\ds}_{\mathrm{out}}$
will have energy $\tilde{\lambda}_{\us G_+ \ds} = - J_2 ( 1 - \hlf \Delta)$
which is very different from the energies 
$\tilde{\lambda}_{\us \ds \ds \us}  = J_2 (1 + \hlf \Delta)+J_1\Delta$
and $\tilde{\lambda}_{\ds \us \us \ds}  =-J_2 (1 + \frac{3}{2} \Delta)+J_1\Delta$
of the states to which it can couple via a single spin-exchange
(assuming $\Delta \neq 0$, see below and the supplementary material), 
see Fig.~\ref{fig:spchTr}(b1). Therefore, such an initial state will 
remain stationary and the control spin excitation on the gate 
will block the transfer of the target spin between the input and output 
ports, see Fig.~\ref{fig:spchTr}(c). Exactly the same arguments apply 
to the initial state  $\ket{\us G_{-} \ds}$ with the magnetic
field set to $h=h_+$.

In the same spirit, we can construct spin transistors with longer spin 
chains (see the supplementary material for $N=5$), the above case of $N=4$ being 
the shortest and simplest one. The general idea illustrated
in Fig.~\ref{fig:spchTr}(a-b) is as follows:
The gate region consists of $N-2$ strongly-coupled spins, 
$J_j \gg J_{1,N-1}$ for all $j \in [2,N-2]$. Therefore, the single excitation 
space of the gate has $N-2$ eigenstates $\ket{G_{i}}$ 
split by $\delta \lambda_G \sim \frac{4J_j}{N-2}$. 
With the magnetic field $h_j$, we tune one of these eigenstates,
say $\ket{G_{i'}}$, in resonance with the single-excitation input
$\ket{\us \ds \ldots \ds \ds}$ and output  $\ket{\ds \ds \ldots \ds \us}$
states. Assuming $J_{1,N-1} \ll \delta \lambda_G$, all the other eigenstates  
$\ket{G_{i}}$ will remain decoupled during the transfer, and we will 
have a simple three-level dynamics for a single target spin.
To close the gate, we place a single spin excitation in one of the 
gate eigenstates $\ket{G_{i \neq i'}}$ from where the control spin cannot
leak out since this eigenstate is non-resonant. Simultaneously, the target 
spin cannot enter the gate region since the double-excitation subspace,
to which it is coupled, is shifted in energy due to the spin-spin interaction,
resulting in the transfer blockade.

\section*{Physical realization}
A possible system to realize the spin chain Hamiltonian in Eq.~(\ref{eg:ham}) is 
a cold ensemble of $N$ strongly-interacting atoms in a one-dimensional 
trap \cite{volosniev2014,volosniev2015,deuretzbacher2014,levinsen2015}. 
Recent experiments have confirmed that spin chains may indeed be
realized this way \cite{murmann2015}.
An outline of the procedure to map this system onto the 
Heisenberg $XXZ$ spin model is presented in the supplementary materials.
A pair of internal atomic states can serve as the spin-up and 
spin-down states. In the full model, the strong contact interactions 
between the atoms are described by the dimensionless coefficients 
$g_{\us \ds} \equiv g \gg 1$ and $g_{\us \us} = g_{\ds \ds} = \kappa g$,
where the parameter $\kappa > 0$ is related to the asymmetry parameter 
of the effective Heisenberg spin-$\hlf$ model as
$\Delta = \left(1 - \frac{2}{\kappa} \right)$. In turn, the exchange
constants of the Heisenberg model $J_j = - \frac{\alpha_j}{g}$ 
are proportional to the geometric factors $\alpha_j$ which are 
determined by the single-particle solutions of the Schr\"odinger 
equation in a one-dimensional confining potential $V(x)$. Hence, 
the shape of the trapping potential can be used to tune the
necessary parameters of the effective spin chain~\cite{volosniev2015}.

\begin{figure}
\centering
\includegraphics[scale=0.7]{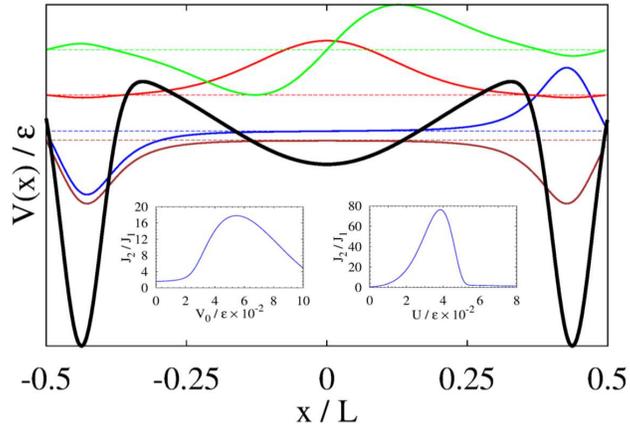}
 \caption{Shape of the potential in Eq.~(\ref{eq:1Dpot}) 
(thick solid black curve), and single-particle eigenfunctions
(thinner solid curves) corresponding to the four lowest-energy levels 
(dashed horizontal lines), for $V_0 = 500 \varepsilon$ and $U = 200 \varepsilon$, where $\varepsilon = \frac{1}{mL^2}$ with $m$ the atom mass.
The insets show the ratio of exchange coefficients $J_2/J_1$ 
as a function of $V_0$, for $U = 200 \varepsilon$ (left), and
as a function of $U$, for $V_0 = 500 \varepsilon$ (right). 
Other parameters in Eq.~(\ref{eq:1Dpot})  
are $a = \frac{384}{L^2}$, $b = \frac{64}{5L^2}$,
$x_0 = \frac{7 L}{16}$.}
\label{fig:1Dpot}
\end{figure}

To realize the Hamiltonian~(\ref{eg:ham}) for $N=4$
particles with $J_{1,3} / J_2 \ll 1$, one may use a triple-well
potential. This may be modeled in numerous ways, and we choose 
the simple form
\begin{equation}
\label{eq:1Dpot}
V(x) = -V_0 [ e^{-a (x-x_0)^2} + e^{-a (x+x_0)^2}] - U e^{-b x^2} ,
\end{equation}
shown in Fig.~\ref{fig:1Dpot} (see the caption for the parameters 
$a,b$ and $x_{0}$). The potential consists of 
a shallow Gaussian well at the center and a pair of deeper and
narrower wells next to the boundaries. The four lowest-energy
single particle wavefunctions of the potential~\eqref{eq:1Dpot}
are also shown in Fig.~\ref{fig:1Dpot}. The two lower energy states
are nearly degenerate and the corresponding wavefunctions have 
sizable amplitudes at the deep wells near the boundaries, while the
two higher energy states have much larger energy separation, with
the amplitudes of the corresponding wavefunctions being large in 
the shallow well in the middle. Accordingly, the 
effective exchange interactions $J_{1}=J_{3}$ is much smaller than $J_2$. 
The dependence of the ratio $J_2/J_1$ on the parameters $V_0$ and $U$ in 
Eq.~\eqref{eq:1Dpot} are shown in the insets of Fig.~\ref{fig:1Dpot}.  

\begin{figure}
\centering
\includegraphics[scale=0.4]{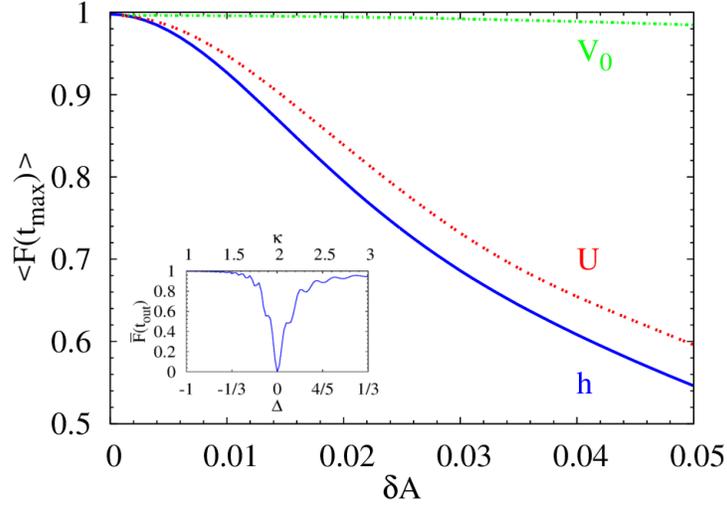}
\caption{The fidelity of transfer $F(t_{\mathrm{out}})$ averaged 
over $100$ independent realizations of the spin chain. 
In each realization, one of the parameters, $A = \{V_0, U, h\}$, 
is a random variable with the Gaussian probability distribution 
around the ideal mean $A_0$ ($V_0 = 500$, $U = 200$, $h = h_{+}$),
with the other two parameters kept constant. 
The standard deviation is $A_0 \delta A$. 
The inset shows the blockade fidelity $\bar{F}(t_{\mathrm{out}})$ for different values
of $\kappa$ (top horizontal axis) or $\Delta$ (bottom horizontal axis).}
\label{fig:Fidels}
\end{figure}

The nearly perfect transfer, or complete blockade, of the target spin
for the open, or closed, gate as shown in Fig.~\ref{fig:spchTr}(c) 
was obtained for the system parameters corresponding to the main panel 
of Fig.~\ref{fig:1Dpot}. It is however important to quantify the sensitivity
of the spin transistor to uncontrolled fluctuations of the parameters.
In Fig.~\ref{fig:Fidels} we show the fidelities $F(t_{\mathrm{out}})
= |\bra{\ds \ds \ds \us} e^{-iHt_{\mathrm{out}}} \ket{\us \ds\ds\ds}|^2$ 
of transfer at time $t_{\mathrm{out}}=\pi/J_1$ versus the amplitude of random 
noise affecting the trapping potential or the magnetic field.
We observe that coherent transfer is quite robust with respect 
to moderate variations in $V_0$, but is rather sensitive to small
variations in $U$ and $h$ since they detrimentally affect the gate 
resonant conditions. In the meantime, the gate blockade is virtually 
unaffected by uncertainties in $U,V_0,h$. In the inset 
of Fig.~\ref{fig:1Dpot} we also show the dependence of blockade fidelity 
$\bar{F}(t_{\mathrm{out}}) = |\bra{\us G_- \ds} e^{-iHt_{\mathrm{out}}} \ket{\us G_- \ds}|^2$ 
on $\kappa$ which determines the asymmetry parameter $\Delta$ of the $XXZ$ 
model. Clearly, the spin transistor cannot operate when $\Delta = 0$,
i.e., in the absence of the spin-spin $\sigma_z^j \sigma_z^{j+1}$ interactions, 
since then the control spin can leak out of the gate, as mentioned above 
and discussed in the supplementary material. 

\section*{Summary and outlook}
To summarize, we have presented a scheme for a quantum spin transistor
realized in a Heisenberg spin chain and proposed and analyzed its physical
implementation with cold trapped atoms. In our scheme, the presence 
$\ket{1}_{\mathrm{gate}}$ or absence $\ket{0}_{\mathrm{gate}}$ of a control 
spin excitation at the gate can block or allow the transfer of an 
arbitrary target spin state between the input and output ports. 
If the gate is prepared in a superposition of open and closed states,
then the initial state of the system with the target spin-up at the input
port will evolve at time $t_{\mathrm{out}}$ into the spatially entangled state, 
\begin{eqnarray*}
& &\ket{\us}_{\mathrm{in}} \frac{1}{\sqrt{2}} (\ket{0}_{\mathrm{gate}} 
+ \ket{1}_{\mathrm{gate}}) \ket{\ds}_{\mathrm{out}} \\ 
& & \to \frac{1}{\sqrt{2}} 
(\ket{\ds}_{\mathrm{in}} \ket{0}_{\mathrm{gate}} \ket{\us}_{\mathrm{out}} +
\ket{\us}_{\mathrm{in}} \ket{1}_{\mathrm{gate}} \ket{\ds}_{\mathrm{out}}) . 
\end{eqnarray*}
Furthermore, if the gate is integrated into a larger system in which
the excited spins from the ``source'' can be fed (one-by-one or
one after the other) into the input port, and the output port is 
connected to the initially unexcited ``drain'', then the initial 
gate superposition state will result in a (macroscopically) 
entangled Schr\"odinger cat like state of many spins,
\begin{eqnarray*}
& & \ket{\us \us \ldots \us }_{\mathrm{in}} \frac{1}{\sqrt{2}} 
(\ket{0}_{\mathrm{gate}} + \ket{1}_{\mathrm{gate}}) 
\ket{\ds \ds \ldots \ds}_{\mathrm{out}} \\
& & \to \frac{1}{\sqrt{2}} (\ket{\ds \ds \ldots \ds }_{\mathrm{in}} 
\ket{0}_{\mathrm{gate}} \ket{\us \us \ldots \us }_{\mathrm{out}} \\
& & \qquad \; + \ket{\us \us \ldots \us }_{\mathrm{in}} \ket{1}_{\mathrm{gate}}
\ket{\ds \ds \ldots \ds}_{\mathrm{out}}). 
\end{eqnarray*}
We note that with the atomic realization of spin chains, 
with the spin-up and spin-down states corresponding to 
the hyperfine (Zeeman) sublevels of the ground electronic state, 
the preparation of coherent superposition of gate states 
$\ket{0}_{\mathrm{gate}} \equiv \ket{\ds \ds}$ and
$\ket{1}_{\mathrm{gate}} \equiv \ket{G_+}$ (assuming $h=h_-$) 
can be accomplished by applying to $\ket{\ds \ds}$ a microwave 
or two-photon (Raman) optical $\pi/2$-pulse of proper frequency 
to match the energy difference $\delta \lambda = 2J_2 \Delta$ between 
$\ket{\ds \ds}$ and $\ket{G_+}$.

\begin{addendum}
 \item This work was funded in part by the
Danish Council for Independent Research DFF Natural
Sciences and the DFF Sapere Aude program, and by the Carlsberg Foundation.  
A.G.V. acknowledges partial support by Helmholtz Association
under contract HA216/EMMI. 
D.P. is grateful to the Aarhus Institute of Advanced Studies in Denmark 
for hospitality and support, and to the Alexander von Humboldt Foundation 
for support during his stay in Germany. 
 \item[Author Contributions] O.~V.~M., D.~P., A.~G.~V., M.~V. and N.~T.~Z. devised the project. O.~V.~M. and A.~G.~V. developed the 
formalism under the supervision of D.~P. and N.~T.~Z. The numerical calculations were carried out by O.~V.~M. The initial draft of the paper was written by 
O.~V.~M., N.~T.~Z., and D.~P. All authors contributed to the revisions that led to the final version.
 \item[Competing Interests] The authors declare that they have no
competing financial interests.
 \item[Correspondence] Correspondence and requests for materials
should be addressed to D.~P. (email: dap@iesl.forth.gr) or N.~T.~Z (email: zinner@phys.au.dk).
\end{addendum}


\section*{Supplementary Materials}

Here we provide details of calculations for the values of gate 
magnetic field $h_j$ required for the realization of quantum spin 
transistor in a Heisenberg $XXZ$ spin chain described by 
Hamiltonian~(1) with $h_{1,_{N}} = 0$ and $J_i=J_{N-i}$.

\subsection{$N=3$ spin chain.}\label{appendixN=3}
We start with the spin chain of one spin-up and two spin-down particles. 
Our first goal is to transfer the population of the initial state
$\ket{\us \ds \ds}$ to the final state
$\ket{\ds \ds \us }$ via the intermediate state
$\ket{\ds \us \ds}$. 

In the basis of $\{ \ket{\us\ds\ds},
\ket{\ds\us\ds}, \ket{\ds\ds\us} \}$,
Hamiltonian (1) can be written in matrix form as
\begin{equation}
\label{mHam}
H = 
 \begin{pmatrix}
  -h & -J_1 & 0 \\
  -J_1 & h + J_1 \Delta & -J_1\\
  0 & -J_1 & -h
 \end{pmatrix},
\end{equation}
where $h \equiv h_2$. The eigenvalues of $H$ read
\begin{subequations}
\label{evals1+2}
\begin{align}
 &\lambda_1 = \frac{1}{2} J_1  \Delta - \sqrt{2J_1^2 + \left ( h + \frac{1}{2}J_1 \Delta \right )^2}, \\
 &\lambda_2 = -h,\\
 &\lambda_3 = \frac{1}{2} J_1  \Delta + \sqrt{2J_1^2 +  \left ( h + \frac{1}{2}J_1 \Delta \right )^2},
\end{align}
\end{subequations}
and the corresponding non-normalized eigenvectors are
\begin{subequations}
\label{evecs1+2}
\begin{align}
 &\ket{\Psi_1} = \{1,  - \frac{h+\lambda_1}{J_1}, 1 \}, \\
 &\ket{\Psi_2} = \{1, 0, -1\},\\
 &\ket{\Psi_3} = \{1,  - \frac{h+\lambda_3}{J_1}, 1 \}.
\end{align}
\end{subequations}

We expand the initial and final states in the basis of eigenvectors 
of Eq.~\eqref{evecs1+2} as 
\begin{subequations}
\label{eq:expan3}
\begin{align}
 &\ket{\us\ds\ds} = \sum_{k=1}^N a^{(1)}_k \ket{\Psi_k}, \\
 &\ket{\ds\ds\us} = \sum_{k=1}^N a^{(3)}_k \ket{\Psi_k} ,
\end{align}
\end{subequations}
where $a^{(i)}_k$ are the expansion coefficients, the index $i = 1, 2, 3$ 
denotes the expanded spin state. Since the Hamiltonian is a bisymmetric 
matrix, the coefficients $a^{(1)}_k$ and $a^{(3)}_k$ are related as 
$a^{(1)}_k = (-1)^{k+1} a^{(3)}_k$ \cite{tao2002,nield1994}. 
Therefore, if we apply the evolution operator $U(t) = e^{-iH t}$
to the initial state
and expand the spin states, we obtain the condition for perfect state transfer after $t_{out}$, $U(t_{\mathrm{out}}) \ket{\us\ds\ds} = \ket{\ds\ds\us}$, in the following form
\begin{equation}
 \sum_{k=1}^N [ e^{-i\lambda_k t_{\mathrm{out}}} - (-1)^{k+1}  ] 
a^{(1)}_k \ket{\Psi_k} = 0.
\end{equation}
Apparently, the conditions for perfect excitation transfer in this spin chain 
after the time interval $t_{\mathrm{out}}$ are
\begin{subequations}
 \begin{align}
  &(\lambda_2 - \lambda_1) t_{\mathrm{out}} = (2 m_1 + 1) \pi, \\
  &(\lambda_3 - \lambda_2) t_{\mathrm{out}} = (2 m_2 - 1) \pi,
 \end{align}
\end{subequations}
where $m_1$ and $m_2$ are non-negative integers, while $m_2 > 0$ since we 
assume that $\lambda_1 < \lambda_2 < \lambda_3$. From these conditions,
we can determine the value of magnetic field $h$ for the transfer to occur 
during the shortest possible time, $\min t_{\mathrm{out}} \equiv t_{\mathrm{min}}$. 
We obtain $m_1 = 0$, $m_2 = 1$ and 
$\frac{\lambda_2 - \lambda_1}{\lambda_3 - \lambda_2} = 1$.
We then find the corresponding magnetic field
\begin{equation}
 h = -\frac{1}{2} J_1 \Delta. \label{sup:eq:hN3}
\end{equation}
The minimal time interval for transfer is 
$t_{\mathrm{min}} = \frac{\pi}{\lambda_2 - \lambda_1} =  \frac{\pi}{\sqrt{2}J_1}$.
Note that the energies of the initial $\ket{\us\ds\ds}$ and final
$\ket{\ds\ds\us}$ states, $\frac{1}{2} J_1 \Delta$, are equal to 
the energy of the intermediate state $\ket{\ds\us\ds}$, so perfect 
transfer for a three-level system is achieved when the intermediate 
state is resonant with the initial and final states. This is of course
an obvious result, but its derivation can be useful for examining
more complicated cases with $N>3$.

Consider now the spin chain with two excitations. 
In the basis of $\{ \ket{\us\us\ds},
\ket{\us\ds\us}, \ket{\ds\us\us} \}$,
the Hamiltonian matrix reads
\begin{equation}
H = 
 \begin{pmatrix}
  h & -J_1 & 0 \\
  -J_1 & - h + J_1 \Delta & -J_1\\
  0 & -J_1 & h
 \end{pmatrix},
\end{equation}
Our goal is that the initial state $\ket{\us\us\ds}$ does not evolve in time, 
i.e., the control spin at the $j=2$ site does not leak out to the site $j=3$
while blocking the transfer of spin excitation from the site $j=1$.
We thus require that the initial state $\ket{\us\us\ds}$
(and state $\ket{\ds\us\us}$) be out of resonance with state 
$\ket{\us\ds\us}$. This leads to the condition $J_1 \ll |J_1 \Delta - 2h|$,
which, for the value of the magnetic field as in Eq.~\eqref{sup:eq:hN3},
reduces to $|2 \Delta| \gg 1$. This is the Ising limit of our spin-chain 
Hamiltonian, which is a rather trivial and impractical case, as it would also
require a large magnetic field $|h| \gg J_1$. 

Note finally that in order to use the spin transistor for various quantum 
information tasks described in the main text, the spin excitation on 
the gate site(s) should not leak out to the output or input ports, 
even when alone. Obviously, in the spin chain with $N=3$ and the 
magnetic field as in Eq.~\eqref{sup:eq:hN3}, such an initial state 
$\ket{\ds\us\ds}$ resonantly couples to states $\ket{\us\ds\ds}$ 
and $\ket{\ds\ds\us}$, and therefore the three-site system is not suitable 
for our purposes. Below we show, that for spin chains with $N \geq 4$ 
it is possible to realize a quantum spin transistor for any value of $\Delta$,
except the very special case of $\Delta = 0$ corresponding to the $XX$ model.

\subsection{$N=4$ spin chain.}\label{appendixN=4}

Now we consider the $N_\us = 1$ and $N_\ds = 3$ spin chain.
For a spatially symmetric chain, we have two different interaction 
coefficients, $J_1 (= J_3)$ and $J_2$, and we take $h_{2,3} = h$ ($h_{1,4} =0$).

In the basis of $\left \{ \ket{\us \ds \ds \ds}, \ket{ \ds \us  \ds \ds},
\ket{\ds\ds\us\ds}, \ket{\ds\ds\ds\us} \right \}$, the Hamiltonian matrix is
\begin{equation}
\label{mHam1+3}
H = 
 \begin{pmatrix}
  -2h - \frac{1}{2}J_2 \Delta & -J_1 & 0 & 0 \\
  -J_1 & \frac{1}{2}J_2 \Delta & -J_2 & 0\\
  0 & -J_2 & \frac{1}{2}J_2 \Delta & -J_1\\
  0 & 0 & -J_1 & -2h - \frac{1}{2}J_2 \Delta  
 \end{pmatrix}.
\end{equation}
We assume that the strongest interaction $J_2 \gg J_1$ is between states 
$\ket{ \ds \us  \ds \ds}$ and $\ket{\ds\ds\us\ds}$. 
To utilize this fact, we prediagonalize the Hamiltonian
by finding the eigenvalues and eigenvectors of the inner (gate) block 
$\begin{pmatrix} \frac{1}{2}J_2 \Delta & -J_2 \\
-J_2 & \frac{1}{2}J_2 \Delta \end{pmatrix}$, which are given by 
$\lambda_+ = \frac{1}{2}J_2(\Delta - 2)$,
$\lambda_- = \frac{1}{2}J_2(\Delta + 2)$, 
and $\ket{G_+} = \frac{1}{\sqrt{2}}(\ket{\us\ds} + \ket{\ds\us})$, 
$\ket{G_-} = \frac{1}{\sqrt{2}}(\ket{\us\ds} - \ket{\ds\us})$.
Our new basis is then $\left \{ \ket{\us \ds \ds \ds}, 
\frac{1}{\sqrt{2}}\left ( \ket{ \ds \us  \ds \ds} + \ket{ \ds \ds  \us \ds} \right ),
\frac{1}{\sqrt{2}}\left ( \ket{ \ds \us  \ds \ds} - \ket{ \ds \ds  \us \ds} \right ),
\ket{\ds\ds\ds\us} \right \}$. We introduce a unitary transformation
\begin{equation}
\label{eq:U1+3}
O = 
 \begin{pmatrix}
  1 & 0 & 0 & 0 \\
  0 & \frac{1}{\sqrt{2}} & \frac{1}{\sqrt{2}} & 0\\
  0 & \frac{1}{\sqrt{2}} & -\frac{1}{\sqrt{2}} & 0\\
  0 & 0 & 0 & 1  
 \end{pmatrix},
\end{equation}
which, when applied to Hamiltonian~(\ref{mHam1+3}) leads to
\begin{equation}
\label{eq:trHam1+3}
\tilde{H} = 
 \begin{pmatrix}
  -2h - \frac{1}{2}J_2 \Delta & -\frac{1}{\sqrt{2}}J_1 & -\frac{1}{\sqrt{2}}J_1 & 0 \\
  -\frac{1}{\sqrt{2}}J_1 & \frac{1}{2}J_2\Delta - J_2 & 0 & -\frac{1}{\sqrt{2}}J_1\\
  -\frac{1}{\sqrt{2}}J_1 & 0 & \frac{1}{2}J_2\Delta + J_2 & \frac{1}{\sqrt{2}}J_1\\
  0 & -\frac{1}{\sqrt{2}}J_1 & \frac{1}{\sqrt{2}}J_1 & -2h - \frac{1}{2}J_2 \Delta 
 \end{pmatrix}.
\end{equation}

We can now find magnetic fields $h$ needed to achieve the spin transfer. 
Note that in the limit of $J_1 \to 0$ Hamiltonian~\eqref{eq:trHam1+3} is 
diagonal, with the energy levels
\begin{subequations}
\label{evalsTransf1+3}
\begin{align}
 &\tilde{\lambda}_1 = -2h - \frac{1}{2}J_2 \Delta  , \\
 &\tilde{\lambda}_2 = \frac{1}{2} J_2 (\Delta - 2 ), \\
 &\tilde{\lambda}_3 = \frac{1}{2} J_2 (\Delta + 2 ), \\
 &\tilde{\lambda}_4 =  -2h - \frac{1}{2}J_2 \Delta.
\end{align}
\end{subequations}
Spin excitation transfer can be obtained if there are three levels in resonance with each other, 
i.e., either $\tilde{\lambda}_{1,4} = \tilde{\lambda}_2 (=\lambda_+)$ or 
$\tilde{\lambda}_{1,4} = \tilde{\lambda}_3(=\lambda_-)$.
The two values of the magnetic field that satisfy these conditions are
\begin{subequations}
\label{hpm3+1}
\begin{align}
& h_{+} = \frac{1}{2} J_2 (1 - \Delta), \\
& h_{-} = -\frac{1}{2} J_2 (1 + \Delta).
\end{align}
\end{subequations}
Hence, for any $\Delta$ it is possible to reduce this spin chain to an effective three-state system.

To verify that (nearly) perfect transfer is indeed achieved, we use the same approach as 
for the three-spin case. The eigenvalues of Hamiltonian~\eqref{mHam1+3} are
\begin{subequations}
\label{evals1+3}
\begin{align}
 &\lambda_1 = -\frac{1}{2} J_2 - h - \sqrt{J_1^2 + \left ( h -  \frac{1}{2} J_2 (1 - \Delta) \right )^2}, \\
 &\lambda_2 = \frac{1}{2} J_2 - h - \sqrt{J_1^2 + \left ( h +\frac{1}{2} J_2 (1 + \Delta) \right )^2}, \\
 &\lambda_3 = -\frac{1}{2} J_2 - h + \sqrt{J_1^2 + \left ( h -  \frac{1}{2} J_2 (1 - \Delta) \right )^2},\\
 &\lambda_4 = \frac{1}{2} J_2 - h + \sqrt{J_1^2 + \left ( h +\frac{1}{2} J_2 (1 + \Delta) \right )^2},
\end{align}
\end{subequations}
and the corresponding non-normalized eigenvectors in the basis of 
$\left \{ \ket{\us \ds \ds \ds}, \ket{ \ds \us  \ds \ds},
\ket{\ds\ds\us\ds}, \ket{\ds\ds\ds\us} \right \}$ are 
\begin{subequations}
\label{evecs1+3}
\begin{align}
&\ket{\Psi_1} = \left \{1,  \frac{ J_1^2 +  J_2((1-\frac{1}{2}\Delta)J_2 + \lambda_3)}{ J_1 \left (\frac{1}{2} J_2 \Delta - \lambda_1 \right )},  \frac{J_2 (1 - \frac{1}{2}\Delta) + \lambda_3}{J_1}, 1 \right \}, \\
&\ket{\Psi_2} = \left \{1,   \frac{ J_1^2 + J_2 ((1 + \frac{1}{2}\Delta)J_2 - \lambda_4)}{J_1 (\frac{1}{2} J_2 \Delta - \lambda_2)},		  
		     \frac{J_2(1+ \frac{1}{2} \Delta) - \lambda_4}{J_1}, -1 \right \}, \\	     
 &\ket{\Psi_3} = \left \{1,  \frac{ J_1^2 + J_2 \left((1-\frac{1}{2}\Delta)J_2 + \lambda_1 \right )}{ J_1\left (\frac{1}{2}J_2\Delta- \lambda_3\right )},
		     \frac{J_2(1 - \frac{1}{2}\Delta)+ \lambda_1}{J_1}, 1 \right \}, \\		
&\ket{\Psi_4} = \left \{1,   \frac{ J_1^2 + J_2 \left( (1+\frac{1}{2}\Delta)J_2 - \lambda_2 \right )}{ J_1\left (\frac{1}{2}J_2\Delta- \lambda_4\right )},	
\frac{J_2(1 + \frac{1}{2}\Delta) - \lambda_2}{J_1}, -1 \right \}.
\end{align}
\end{subequations}

As before, we expand the initial and final states in the basis of eigenvectors from Eq. \eqref{evecs1+3},
\begin{subequations}
 \begin{align}
  \ket{\us \ds \ds \ds} = \sum_{k=1}^N {a^{(1)}_k \ket{\Psi_k}}, \\
  \ket{\ds \ds \ds \us} = \sum_{k=1}^N {a^{(4)}_k \ket{\Psi_k}}.
 \end{align}
\end{subequations}
Since Hamiltonian~\eqref{mHam1+3} is a bisymmetric matrix, the expansion coefficients $a^{(i)}_{k}$ are related as
$a^{(1)}_1 = a^{(4)}_1$, $a^{(1)}_2 = - a^{(4)}_2$, $a^{(1)}_3 = a^{(4)}_3$, $a^{(1)}_4 = - a^{(4)}_4$\cite{tao2002, nield1994}.
Hence, the necessary and sufficient conditions for state $\ket{\us \ds \ds \ds}$ to evolve into
state $\ket{\ds \ds \ds\us }$ during time $t_{\mathrm{out}}$ are
\begin{equation}
 (\lambda_{k+1} - \lambda_{k})t_{\mathrm{out}} = \pi (2 m_k + 1),
\end{equation}
where $m_k$ is a positive integer.

We now use the values of the magnetic field in Eqs.~(\ref{hpm3+1}) to determine 
the fastest transfer time, $t_{\mathrm{min}} = \frac{\pi}{\delta \lambda}$,
where $\delta \lambda$ is the energy difference between the equidistant levels. 
For $h_{+}$ we obtain
\begin{subequations}
 \begin{align}
  &\lambda_2 - \lambda_1 = J_1 + J_2 - \sqrt{J_1^2 + J_2^2} \approx J_1, \\
  &\lambda_3 - \lambda_2 = J_1 - J_2 + \sqrt{J_1^2 + J_2^2} \approx J_1, \\
  &\lambda_4 - \lambda_3 = - J_1 + J_2 + \sqrt{J_1^2 + J_2^2} \approx 2J_2 - J_1.
 \end{align}
\end{subequations}
We see that if $J_1 / J_2 \ll 1$ the lowest three energy levels of the system
are equidistant, with the difference $\delta \lambda = J_1$, and the highest
energy level lies far away from the others. That is why we can expect nearly
perfect transfer at time $t_{\mathrm{min}} = \frac{\pi}{J_1}$. 

For $h_{-}$, we have 
\begin{subequations}
 \begin{align}
  &\lambda_2 - \lambda_1 = J_2 - J_1 + \sqrt{J_1^2 + J_2^2} \approx 2J_2 - J_1 \\
  &\lambda_3 - \lambda_2 = J_1 - J_2 + \sqrt{J_1^2 + J_2^2} \approx J_1 \\
  &\lambda_4 - \lambda_3 = J_1 + J_2 - \sqrt{J_1^2 + J_2^2} \approx J_1
 \end{align}
\end{subequations}
with the higher three levels equidistant and the lower level is
much further below the other three.	

Consider an example with the particular values of $\Delta = -1$ 
and $h_{-} = -\frac{1}{2} J_2 (1 + \Delta) = 0$. 
The normalized eigenstates obtained from Eqs.~\eqref{evecs1+3} are
\begin{subequations}
\begin{align}
 &\ket{\Psi_1} =  \frac{1}{2\sqrt{J_1^2 + J_2^2 + J_2 \sqrt{J_1^2 + J_2^2}}} \left \{ J_1,  J_2  + \sqrt{J_1^2 + J_2^2},
		    J_2  + \sqrt{J_1^2 + J_2^2}, J_1 \right \}, \\
&\ket{\Psi_2} = \frac{1}{2} \left \{1,  1, -1, -1 \right \}, \\
 &\ket{\Psi_3} = \frac{1}{2\sqrt{J_1^2 + J_2^2 - J_2 \sqrt{J_1^2 + J_2^2}}} \left \{ J_1,  J_2  - \sqrt{J_1^2 + J_2^2},
		    J_2  - \sqrt{J_1^2 + J_2^2}, J_1 \right \}, \\		     
 &\ket{\Psi_4} = \frac{1}{2} \left \{1,  -1, 1, -1 \right \},
\end{align}
\end{subequations}
and the spin states expanded in this basis are
\begin{subequations}
\begin{align}
 &\ket{\us \ds \ds \ds} =  \frac{1}{2} \left \{  \frac{J_1}{\sqrt{J_1^2 + J_2^2 + J_2 \sqrt{J_1^2 + J_2^2}}},  1,
		    \frac{J_1}{\sqrt{J_1^2 + J_2^2 - J_2 \sqrt{J_1^2 + J_2^2}}}, 1 \right \}_{\Psi}, \\
&\ket{\ds \us \ds \ds} = \frac{1}{2} \left \{  \frac{J_2  + \sqrt{J_1^2 + J_2^2}}{\sqrt{J_1^2 + J_2^2 + J_2 \sqrt{J_1^2 + J_2^2}}},  
		    1, \frac{J_2  - \sqrt{J_1^2 + J_2^2}}{\sqrt{J_1^2 + J_2^2 - J_2 \sqrt{J_1^2 + J_2^2}}}, -1 \right \}_{\Psi}, \\
 &\ket{\ds \ds \us \ds} = \frac{1}{2} \left \{  \frac{J_2  + \sqrt{J_1^2 + J_2^2}}{\sqrt{J_1^2 + J_2^2 + J_2 \sqrt{J_1^2 + J_2^2}}},  
		    -1, \frac{J_2  - \sqrt{J_1^2 + J_2^2}}{\sqrt{J_1^2 + J_2^2 - J_2 \sqrt{J_1^2 + J_2^2}}}, 1 \right \}_{\Psi}, \\		     
 &\ket{\ds \ds \ds \us} = \frac{1}{2} \left \{  \frac{J_1}{\sqrt{J_1^2 + J_2^2 + J_2 \sqrt{J_1^2 + J_2^2}}},  -1,
		    \frac{J_1}{\sqrt{J_1^2 + J_2^2 - J_2 \sqrt{J_1^2 + J_2^2}}}, -1 \right \}_{\Psi}.
\end{align}
\end{subequations}
In the limit of $J_1 / J_2 \ll 1$, these expressions reduce to
\begin{subequations}
\begin{align}
 &\ket{\us \ds \ds \ds} \simeq \frac{1}{2} \left \{  \frac{J_1}{\sqrt{2}J_2},  1, \sqrt{2}, 1 \right \}_{\Psi}, \\
&\ket{\ds \us \ds \ds} \simeq \frac{1}{2} \left \{ \sqrt{2}, 1, -\frac{J_1}{\sqrt{2}J_2}, -1 \right \}_{\Psi}, \\
 &\ket{\ds \ds \us \ds} \simeq \frac{1}{2} \left \{ \sqrt{2}, -1, -\frac{J_1}{\sqrt{2}J_2}, 1 \right \}_{\Psi}, \\
 &\ket{\ds \ds \ds \us} \simeq \frac{1}{2} \left \{  \frac{J_1}{\sqrt{2}J_2},  -1, \sqrt{2}, -1 \right \}_{\Psi}.
\end{align}
\end{subequations}
We see that, for all spin states, one of the coefficients of the expansion in 
the basis of energy eigenstates is much smaller than the other three. Hence, 
we indeed have an effective three-level system for spin transfer between 
the input $j=1$ and output $j=4$ ports via the intermediate resonant 
$\ket{\ds G_- \ds}$ state. 
In the main text, we use the following notation for such input and output states: 
\begin{eqnarray*}
& \ket{\us \ds \ds \ds} \equiv \ket{\us}_{\mathrm{in}} \ket{0}_{\mathrm{gate}} \ket{\ds}_{\mathrm{out}} , \\
& \ket{\ds \ds \ds \us} \equiv \ket{\ds}_{\mathrm{in}} \ket{0}_{\mathrm{gate}} \ket{\us}_{\mathrm{out}} .
\end{eqnarray*}

Note that if we place a spin excitation in the non-resonant $\ket{G_+}$
state of the gate (sites 2 and 3), it will stay there indefinitely 
(assuming no energy relaxations) due to large energy mismatch to the other states; 
such a state in our notation would be
\[
\ket{\ds G_+ \ds} \equiv \ket{\ds}_{\mathrm{in}} \ket{1}_{\mathrm{gate}} \ket{\ds}_{\mathrm{out}} .
\]  

Recall that we set the gate magnetic field to $h_-$. If, instead, we chose $h_+$ then 
the roles of the $\ket{G_-}$ and $\ket{G_+}$ states of the gate would be interchanged, 
i.e., $\ket{\ds  G_+ \ds}$ would be the intermediate resonant state for the spin 
transfer between $\ket{\us}_{\mathrm{in}} \ket{0}_{\mathrm{gate}} \ket{\ds}_{\mathrm{out}}$ 
and $\ket{\ds}_{\mathrm{in}} \ket{0}_{\mathrm{gate}} \ket{\us}_{\mathrm{out}}$, while 
$\ket{\ds  G_- \ds} \equiv \ket{\ds}_{\mathrm{in}} \ket{1}_{\mathrm{gate}} \ket{\ds}_{\mathrm{out}}$
would be the trapped (non-evolving) state.

Consider now the two-excitation case, $N_\us = 2$, $N_\ds = 2$. 
In the basis of $\{\ket{\us\us\ds\ds}$,
$\ket{\us\ds\us\ds}$, $\ket{\us\ds\ds\us}$, $\ket{\ds\us\us\ds}$,
$\ket{\ds\us\ds\us}$, $\ket{\ds\ds\us\us} \}$, the Hamiltonian matrix is 
{\small
\begin{equation}
\label{mHam2+2}
H = 
 \begin{pmatrix}
   (-J_1 + \frac{1}{2}J_2 )\Delta & -J_2 & 0 & 0 & 0 & 0 \\
  -J_2 & (J_1 + \frac{1}{2}J_2)\Delta & -J_1 & -J_1 & 0 & 0\\
  0 & -J_1 & (J_1 - \frac{1}{2}J_2)\Delta - 2h & 0 & -J_1 & 0\\
  0 & -J_1 & 0 & (J_1  - \frac{1}{2}J_2)\Delta + 2h & -J_1 & 0\\
  0 & 0 & -J_1 & -J_1 & (J_1  + \frac{1}{2}J_2)\Delta& -J_2\\ 
  0 & 0 & 0 & 0 & -J_2 & (-J_1 + \frac{1}{2}J_2)\Delta
 \end{pmatrix},
\end{equation}
}
By construction, the exchange interaction $J_2$ between the gate sites $j=2-3$
is the strongest. We again prediagonalize the Hamiltonian and change the basis 
to 
\[
\{ \ket{\us G_+\ds}, \ket{\us G_- \ds}, \ket{\us \ds \ds \us}, \ket{\ds \us \us \ds}, 
\ket{\ds G_+ \us}, \ket{\ds G_- \us} \}
\]
with 
$\ket{G_+} \equiv \frac{1}{\sqrt{2}} (\ket{\us\ds} + \ket{\ds\us} )$ 
and $\ket{G_-} \equiv \frac{1}{\sqrt{2}} (\ket{\us\ds} - \ket{\ds\us} )$
being the two eigenstates of the gate part of the chain.
Using the unitary transformation
\begin{equation}
\label{eq:U2+2}
O = 
 \begin{pmatrix}
  \frac{1}{\sqrt{2}} & \frac{1}{\sqrt{2}} & 0 & 0 & 0 & 0 \\
  \frac{1}{\sqrt{2}} & -\frac{1}{\sqrt{2}} & 0 & 0 & 0 & 0 \\
  0 & 0 & 1 & 0 & 0 & 0 \\
  0 & 0 & 0 & 1 & 0 & 0 \\  
  0 & 0 & 0 & 0 & \frac{1}{\sqrt{2}} & \frac{1}{\sqrt{2}}\\
  0 & 0 & 0 & 0 & \frac{1}{\sqrt{2}} & -\frac{1}{\sqrt{2}}
 \end{pmatrix},
\end{equation}
Hamiltonian~\eqref{mHam2+2} transforms into
\begin{equation}
\label{mHam2+2_diag}
\tilde H = 
 \begin{pmatrix}
   -J_2(1 - \frac{1}{2}\Delta)  & -J_1 \Delta & -\frac{1}{\sqrt{2}}J_1 & -\frac{1}{\sqrt{2}}J_1 & 0 & 0 \\
  -J_1 \Delta & J_2 (1 + \frac{1}{2}\Delta) & \frac{1}{\sqrt{2}}J_1 & \frac{1}{\sqrt{2}}J_1 & 0 & 0\\
  -\frac{1}{\sqrt{2}}J_1 & \frac{1}{\sqrt{2}}J_1 & (J_1 - \frac{1}{2}J_2)\Delta - 2h & 0 & -\frac{1}{\sqrt{2}}J_1 & -\frac{1}{\sqrt{2}}J_1 \\
  -\frac{1}{\sqrt{2}}J_1 & \frac{1}{\sqrt{2}}J_1 & 0 & (J_1 - \frac{1}{2}J_2)\Delta + 2h & -\frac{1}{\sqrt{2}}J_1 & -\frac{1}{\sqrt{2}}J_1 \\
  0 & 0 & -\frac{1}{\sqrt{2}}J_1 & -\frac{1}{\sqrt{2}}J_1 &  -J_2(1 - \frac{1}{2}\Delta)& J_1 \Delta\\ 
  0 & 0 &  -\frac{1}{\sqrt{2}}J_1 &  -\frac{1}{\sqrt{2}}J_1 & J_1 \Delta & J_2(1 + \frac{1}{2}\Delta)
 \end{pmatrix}.
\end{equation}

We place one (control) spin-up in the gate, either in state $\ket{G_+} = \ket{1}_{\mathrm{gate}}$, 
if we set the gate magnetic field to $h_-$, or in state $\ket{G_-} = \ket{1}_{\mathrm{gate}}$, 
if we set it to $h_+$. As stated above, the control spin-up then cannot leak out of the gate, 
i.e., $\ket{\ds}_{\mathrm{in}} \ket{1}_{\mathrm{gate}} \ket{\ds}_{\mathrm{out}}$ is stationary.
Next, we place the target spin-up on site $j=1$, obtaining state 
$\ket{\us}_{\mathrm{in}} \ket{1}_{\mathrm{gate}} \ket{\ds}_{\mathrm{out}}$. 
We now verify that this state does not evolve since it is non-resonant with 
all the other states to which it can couple with rates $\sim J_1 \ll J_2$.

Assuming $h_{-} = -\frac{1}{2} J_2 (1 + \Delta)$, the energy of the initial
state $\ket{\us G_+ \ds}$ is 
\begin{equation}
\tilde{\lambda}_{\us G_+ \ds} = - J_2 \left( 1 - \frac{1}{2} \Delta \right) , 
\end{equation}
while the energies of states $\ket{\us \ds \ds \us}$ and $\ket{\ds \us \us \ds}$ are 
\begin{subequations}
\begin{align}
& \tilde{\lambda}_{\us \ds \ds \us} \simeq - \frac{1}{2}J_2\Delta - 2h_- = J_2 \left( 1 +\frac{1}{2} \Delta \right) , \\
& \tilde{\lambda}_{\ds \us \us \ds} \simeq - \frac{1}{2}J_2\Delta + 2h_- =-J_2 \left( 1 +\frac{3}{2} \Delta \right) . 
\end{align}
\end{subequations}
Unless $\Delta =0$, the initial state $\ket{\us G_+\ds}$ is highly non-resonant with all the other connected
states (state $\ket{\ds G_+ \us}$ has of course the same energy, but it is not directly connected to 
$\ket{\us G_+ \ds}$)

Similarly, for $h_{+} = \frac{1}{2} J_2 (1 - \Delta)$, the energy of the initial
state $\ket{\us G_- \ds}$ is 
\begin{equation}
\tilde{\lambda}_{\us G_- \ds} = J_2 \left( 1 + \frac{1}{2} \Delta \right) , 
\end{equation}
while the energies of states $\ket{\us \ds \ds \us}$ and $\ket{\ds \us \us \ds}$ are 
\begin{subequations}
\begin{align}
& \tilde{\lambda}_{\us \ds \ds \us} \simeq - \frac{1}{2}J_2\Delta - 2h_+ = -J_2 \left( 1 - \frac{1}{2} \Delta \right) , \\
& \tilde{\lambda}_{\ds \us \us \ds} \simeq - \frac{1}{2}J_2\Delta + 2h_+ =  J_2 \left( 1 - \frac{3}{2} \Delta \right) , 
\end{align}
\end{subequations}
and again we have a non-resonant initial state $\ket{\us G_- \ds}$ (and it is not directly connected to
$\ket{\ds G_- \us}$). 

\subsection{$N=5$ spin chain.}\label{appendixN=5}

The five-spin chain does not differ much in principle from the four-spin chain. So in the spirit of previous
sections, we start with looking for values of magnetic field, such that spin excitation transfer is achieved
in a $N_\us = 1$, $N_\ds = 4$ spin chain. 
Since we postulate the symmetry of the confining potential, we have only two values of interaction coefficients: 
$J_1$ and $J_2$, with the same condition $J_1/J_2 \ll 1$. However, the magnetic field in the middle of the chain might
not be the same on different sites, and we therefore assume  $h_1 = h_5 = 0$, $h_2 = h_4 = h'$, and $h_3 = h$.

In the basis of $\left\{\ket{\us\ds\ds\ds\ds}, \ket{\ds\us\ds\ds\ds}, \ket{\ds\ds\us\ds\ds},
\ket{\ds\ds\ds\us\ds}, \ket{\ds\ds\ds\ds\us} \right \}$, the Hamiltonian for this system is
\begin{equation}
\label{mHam1+4}
H = 
 \begin{pmatrix}
  -2h' - h - J_2\Delta & -J_1 & 0 & 0 & 0 \\
  -J_1 & -h & -J_2 & 0 & 0\\
  0 & -J_2 & h - 2h' + (J_2 - J_1)\Delta & -J_2 & 0\\
  0 & 0 & -J_2 & -h & -J_1\\
  0 & 0 & 0 & -J_1 & -2h' - h - J_2 \Delta 
 \end{pmatrix}.
\end{equation}
By assumption, the exchange interaction is the strongest between the states 
$\ket{\ds\us\ds\ds\ds}, \ket{\ds\ds\us\ds\ds}, \ket{\ds\ds\ds\us\ds}$,
so we once again prediagonalize the Hamiltonian by finding the eigenvalues 
and eigenvectors of the gate matrix  
$\begin{pmatrix}
-h & -J_2 & 0\\
-J_2 & h - 2h' + (J_2 - J_1)\Delta & -J_2 \\
0 & -J_2 & -h
\end{pmatrix}$. 
We can write eigenvalues as
\begin{subequations}
\label{evals1+4_diag}
\begin{align}
 &
\lambda_+ = -  h' + \frac{1}{2}(J_2 - J_1)\Delta + \sqrt{2J_2^2 + \left (h -  h' + \frac{1}{2}(J_2 - J_1)\Delta \right )^2}, \\
 &\lambda_0 = -h,\\
 &\lambda_- = -  h' + \frac{1}{2}(J_2 - J_1)\Delta - \sqrt{2J_2^2 + \left (h -  h' + \frac{1}{2}(J_2 - J_1)\Delta \right )^2},
\end{align}
\end{subequations}
and the corresponding normalized eigenvectors in the basis of $\left \{ \ket{\us\ds\ds},
\ket{\ds\us\ds}, \ket{\ds\ds\us} \right \}$ are
\begin{subequations}
\label{evecs1+4_diag}
\begin{align}
 &\ket{G_+} = \frac{1}{\sqrt{2J_2^2 + (h + \lambda_{+})^2}}\{J_2,  - (h + \lambda_{+}),  J_2 \}, \\
 &\ket{G_0} = \frac{1}{\sqrt{2}}\{1, 0, -1\},\\
 &\ket{G_-} = \frac{1}{\sqrt{2J_2^2 + (h + \lambda_{-})^2}}\{J_2,  - (h + \lambda_{-}),  J_2 \}.
\end{align}
\end{subequations}
We introduce a new basis $\{ \ket{\us \ds \ds \ds \ds},
\ket{\ds G_{+} \ds}, \ket{\ds G_{0} \ds}, \ket{\ds G_- \ds}, \ket{\ds\ds\ds\ds\us} \}$ 
and the corresponding unitary transformation
\begin{equation}
\label{eq:U1+4}
U = 
 \begin{pmatrix}
  1 & 0 & 0 & 0 & 0 \\
  0 & \frac{J_2}{\sqrt{2J_2^2 + (h + \lambda_{+})^2}} & - \frac{h + \lambda_+}{\sqrt{2J_2^2 + (h + \lambda_{+})^2}} &
  \frac{J_2}{\sqrt{2J_2^2 + (h + \lambda_{+})^2}} & 0\\
  0 & \frac{1}{\sqrt{2}} & 0 & -\frac{1}{\sqrt{2}} & 0\\
  0 & \frac{J_2}{\sqrt{2J_2^2 + (h + \lambda_{-})^2}} & - \frac{h + \lambda_-}{\sqrt{2J_2^2 + (h + \lambda_{-})^2}} &
  \frac{J_2}{\sqrt{2J_2^2 + (h + \lambda_{-})^2}} & 0\\
  0 & 0 & 0 & 0 & 1 
 \end{pmatrix}.
\end{equation}
As before, the transformed Hamiltonian is nearly diagonal when $J_1 / J_2 \ll 1$,
\begin{equation}
\label{mHam1+4_diag}
\tilde{H} \simeq 
 \begin{pmatrix}
  -2h' - h - J_2 \Delta & 0 & 0 & 0 & 0 \\
  0 & \lambda_+ & 0 & 0 & 0\\
  0 & 0 & \lambda_0 & 0 & 0\\
  0 & 0 & 0 & \lambda_- & 0\\
  0 & 0 & 0 & 0 & -2h' - h - J_2 \Delta
 \end{pmatrix}.
\end{equation}
To achieve resonant transfer between states $\ket{\us \ds \ds \ds \ds}$ and $\ket{\ds\ds\ds\ds\us}$,
we need to find the values of magnetic field so that
one of $\lambda_{+}, \lambda_{-}$ or $\lambda_0$ is equal to $-2h' - h - J_2 \Delta$.
For instance, from the condition $-2h' - h - J_2\Delta = \lambda_0$ we find $h' = - \frac{1}{2}J_2\Delta$ 
and any $h$. In this case, the spin transfer goes via the resonant intermediate state $\ket{\ds G_{0} \ds}$.
The condition  $-2h' - h - J_2\Delta = \lambda_{+}$ can be satisfied if
\begin{equation}
h' = \frac{J_2^2}{2(h + J_2 \Delta)} - \frac{1}{2} J_2 \Delta  \;\; \& \;\; h < -J_2 \Delta.
\end{equation}
Alternatively, the condition $-2h' - h - J_2\Delta = \lambda_{-}$ give  
\begin{equation}
h' = \frac{J_2^2}{2(h + J_2 \Delta)} - \frac{1}{2} J_2 \Delta  \;\; \& \;\; h > -J_2 \Delta.
\end{equation}
Therefore, depending on the value of $h + J_2 \Delta$, the spin transfer goes via 
one of the states $\ket{\ds G_{\pm} \ds}$. 

Consider now $N_\us = 2$ spin excitations. 
In the basis of states 
\[
\{ \ket{\us \us \ds \ds \ds}, 
\ket{\us \ds \us \ds \ds}, \ket{\us \ds \ds \us \ds }, 
\ket{\us \ds \ds \ds \us}, \ket{\ds \us \us \ds \ds}, 
\ket{\ds \us \ds \us \ds}, \ket{\ds \us \ds \ds \us }, 
\ket{\ds \ds \us \us \ds}, \ket{\ds \ds \us \ds \us},
\ket{\ds \ds \ds \us \us } \}
\]
the Hamiltonian reads 
{\tiny
\begin{equation}
\label{mHam2+3}
H = -h {\mathbb I} +
 \begin{pmatrix}
  -J_1 \Delta & -J_2 & 0 & 0 & 0 & 0 & 0 & 0 & 0 & 0\\
  -J_2 & 2h-2h'+J_2 \Delta & -J_2 & 0 & -J_1& 0 & 0 & 0 & 0 & 0 \\
  0 & -J_2 & J_1 \Delta & -J_1 & 0 & -J_1 & 0 & 0 & 0 & 0\\
  0 & 0 & -J_1 & -2h'+ (J_1 -J_2) \Delta & 0 & 0 & -J_1 & 0 & 0 & 0\\
  0 & -J_1 & 0 & 0 & 2h & -J_2 & 0 & 0 & 0 & 0\\
  0 & 0 & -J_1 & 0 & -J_2 & 2h'+ (J_1 + J_2) \Delta & -J_1 & -J_2 & 0 & 0\\
  0 & 0 & 0 & -J_1 & 0 & -J_1 & J_1 \Delta & 0 & -J_2 & 0\\
  0 & 0 & 0 & 0 & 0 & -J_2 & 0 & 2h & -J_1 & 0\\
  0 & 0 & 0 & 0 & 0 & 0 & -J_2 & -J_1 & 2h-2h' + J_2 \Delta & -J_2\\
  0 & 0 & 0 & 0 & 0 & 0 & 0 & 0 & -J_2 & -J_1 \Delta
 \end{pmatrix},
\end{equation}
}
where ${\mathbb I}$ is the identity matrix.
We prediagonalize the gate part to obtain the new basis
\[
\{ \ket{\us G_+ \ds}, \ket{\us G_0 \ds}, \ket{\us G_- \ds}, \ket{\us \ds \ds \ds \us}, 
\ket{\ds \bar{G}_- \ds}, \ket{\ds \bar{G}_0 \ds}, \ket{\ds \bar{G}_+ \ds},
 \ket{\ds G_- \us}, \ket{\ds G_0 \us}, \ket{\ds G_+ \us}\}
\], 
where $\bar{G}_{\pm,0}$ 
are obtained from $G_{\pm,0}$ via the replacement $\us \leftrightarrow \ds$.
We place the control spin excitation at the gate in one of the states
$\ket{G_{\pm,0}}$ which is different from the one we chose via the magnetic
field above for the resonant transfer. This spin then remains stationary 
since it cannot leave the gate due to the energy mismatch $\sim J_2 \gg J_1$.
It also blocks the resonant transfer of the target spin from the site $j=1$ to site 
$j=5$, which cannot overcome the energy mismatch to the double excitation
states $\ket{\bar{G}_{\pm,0}}$ of the gate.

Similar arguments can be used to construct a spin transistor with longer chains,
as outlined in the main text.

\subsection{Implementation of the $XXZ$ spin chain with strongly interacting atoms.}

Here we outline the procedure to map a system of strongly interacting 
atoms confined in a 1D trapping potential onto the Heisenberg $XXZ$ spin model,
as was recently shown in \cite{volosniev2014,volosniev2015,deuretzbacher2014,levinsen2015}. 

We consider a two-component Bose gas of atoms. Denoting the components
as spin-up and spin-down, the total number of atoms is $N = N_\us + N_\ds$. 
The strong contact interaction between the atoms is modeled by the Dirac 
delta function potential and the Hamiltonian can be written as ($\hbar = 1$)
\begin{equation}
 H_{atom} = \sum_{\sigma = \uparrow, \downarrow} \sum_{j = 1}^{N_\sigma} \left [ H_0(x_{\sigma, j})
   + \frac{g_{\sigma\sigma}}{mL} \sum_{j'>j}^{N_\sigma} \delta(x_{\sigma,j} - x_{\sigma, j'}) \right ]
   + \frac{g_{\uparrow\downarrow}}{mL}\sum_{i=1}^{N_{\uparrow}} \sum_{j'=1}^{N_{\downarrow}}
   \delta(x_{\uparrow,j} - x_{\downarrow, j'}) + \frac{1}{g}\sum_{\sigma = \uparrow, \downarrow} \sum_{j = 1}^{N_\sigma} B(x_{\sigma,j}) \sigma_z^j, 
\label{hamiltonian0}
\end{equation}
where $H_0(x) = -\frac{1}{2m}\frac{\partial^2}{\partial x^2} +
\frac{1}{mL^2} V(x/L)$ is the single-particle Hamiltonian in a 
one-dimensional trapping potential $V(x/L)$ with a characteristic length $L$, 
$x_{\sigma, j}$ is the coordinate of the $j$th particle with spin 
$\sigma = \{\us, \ds \}$, $m$ is the mass assumed equal for all particles, 
$B(x)$ is a spatially inhomogeneous magnetic field, 
and the Pauli $\sigma_z^j$ operator acts on the spin of the $j$th particle. 
The interaction strengths are 
$g_{\uparrow\downarrow} = g_{\downarrow\uparrow} \equiv g > 0$ and
$g_{\uparrow\uparrow} \equiv \kappa g$, where the parameter $\kappa > 0$
determines the interspecies interaction for bosonic atoms, while 
$\kappa \to \infty$ can be seen as the fermionic limit.

In general, the $N$-particle eigenstate of the system can be written 
as~\cite{volosniev2014,volosniev2015}
\begin{equation}
\label{wf-interacting}
 \Psi = \sum_{k} a_k \theta(x_{P_k(1)}, \dots, x_{P_k(N)}) \Psi_0(x_1, \dots, x_N),
\end{equation}
where the summation runs over all $N!$ permutations $P_k$ of coordinates, 
$a_k \in \mathbb{R} $ are the expansion coefficients which depend on the ordering 
of particles, $\theta(x_1, \dots, x_i, \dots, x_j, \dots, x_N) = 1$
if $x_1 < x_2 < \dots < x_i < \dots < x_j < \dots < x_N$ and zero otherwise. 
$\Psi_0$ is the fully antisymmetrized $N$-particle wavefunction, 
i.e., Slater determinant constructed from the single-particle 
solutions of the Schr{\"o}dinger equation for a particle in potential $V(x)$.
As such, $\Psi_0$ describes non-interacting, or fermionic $1/g = 0$,
$N$-particle system with energy $E_0$. Note that $E_0$ is 
$M(N_\uparrow, N_\downarrow) = N!/(N_\uparrow!N_\downarrow!)$ fold degenerate, since 
the energy does not depend on the particle ordering. For small but finite $1/g$, 
the $N$-particle energy of the interacting system can be written 
in linear order in $1/g$ as \cite{volosniev2014,volosniev2015}
\begin{equation}
 E = E_0 - \frac{1}{g} \frac{\sum_{j=1}^{N-1}(A_j + \frac{2}{\kappa} C_j + \frac{2}{\kappa} D_j) \alpha_j}{\sum_{k=1}^{M(N_\uparrow, N_\downarrow)}a_k^2}+\frac{1}{g}\frac{\sum_{k}a_k^2 \sum_{\sigma = \uparrow, \downarrow} \sum_{j = 1}^{N_\sigma} \langle \theta(x_{P_k(1)}, \dots, x_{P_k(N)}) \Psi_0| B(x_{\sigma,j}) \sigma_z^j|\Psi_0 \rangle}{\braket{\theta \Psi_0}{\Psi_0}\sum_k {a_k^2} },
\end{equation}
where $A_j = \sum_{k=1}^{M(N_\downarrow - 1, N_\uparrow - 1)} (a_{j|k} - b_{j|k})^2$, 
$C_j = \sum_{k=1}^{M(N_\downarrow, N_\uparrow - 2)}c_{j|k}^2$ and 
$D_j= \sum_{k=1}^{M(N_\downarrow - 2, N_\uparrow)} d_{j|k}^2 $. 
Here $a_{j|k}$ denote those coefficients $a_k$ in the 
expansion~\eqref{wf-interacting} for which $x_\uparrow$ is at position $j$ 
followed by $x_\downarrow$ at position $j+1$. 
Similarly, coefficients $b_{j|k}$ correspond to $x_\downarrow$ at position $j$
followed by $x_\uparrow$ at position $j+1$,
coefficients $c_{j|k}$ correspond to $x_\uparrow$ at position $j$ followed 
by $x_\uparrow$ at position $j+1$, and
coefficients $d_{j|k}$ correspond to $x_\downarrow$ at position $j$ followed 
by $x_\downarrow$ at position $j+1$.

The geometric factors $\alpha_j$ depend on both the total number of particles 
and the single-particle solutions of the Schr{\"o}dinger equation for a particle
in potential $V(x)$. An explicit expression for $\alpha_j$ reads
\begin{equation}
 \alpha_j = \frac{1}{m^2} \frac{\int \prod_{i=1}^N \mathrm{d}x_i \theta({x_1, \dots, x_N})\delta(x_1 - x_j)(\partial\Psi_0)^2}
 {\int \prod_{i=1}^N \mathrm{d}x_i\theta(x_1,\dots,x_N)|\Psi_0(x_1,\dots,x_N)|^2},
\end{equation}
where $\partial \Psi_0 = \left( \frac{\partial\Psi_0}{\partial x_1} \right)_{x_1 = x_N}$,
i.e., one first takes the partial derivative of the non-interacting $N$-particle
wave function $\Psi_0$ with respect to $x_1$ and then sets $x_1 = x_N$.

For strong interactions, $g \gg 1$, the Hamiltonian in Eq.~\eqref{hamiltonian0}
can be mapped onto the Heisenberg $XXZ$ spin model Hamiltonian \cite{volosniev2015}, 
cf Eq.~(1) in the main text, with
\begin{align}
& \Delta = 1 - \frac{2}{\kappa} , \\
& J_j \equiv - \frac{\alpha_j}{g} ,\\
& h_j \equiv \frac{\int \prod_{i=1}^N \mathrm{d}x_i\theta(x_1,\dots,x_N)|\Psi_0(x_1,\dots,x_N)|^2 B(x_j)}{\int \prod_{i=1}^N \mathrm{d}x_i\theta(x_1,\dots,x_N)|\Psi_0(x_1,\dots,x_N)|^2 } .
\end{align}
Hence, the shape of the one-dimensional confining potential will determine 
the exchange interaction coefficients $J_j$.

We note that the form of the resulting spin-chain Hamiltonian depends on our choice of the phase of $\theta(x_1,...,x_N)\Psi_0$. For instance, if in Eq. (\ref{wf-interacting}) we choose $|\Psi_0|$ instead of $\Psi_0$, then we obtain the spin-chain Hamiltonian of the form \cite{deuretzbacher2014}
\begin{equation}
H_{|\Psi_0|} = \sum_{j = 1}^{N} h_j \sigma_z^j 
- \frac{1}{2} \sum_{j=1}^{N-1} J_j 
[ -\sigma_x^j \sigma_x^{j+1} - \sigma_y^j \sigma_y^{j+1} 
+  \Delta \sigma_z^j \sigma_z^{j+1} ].
\end{equation}
This Hamiltonian is related to $H$ in Eq.~(1) through the unitary transformation \cite{Takahashi} which rotates the spins at every other site, i.e., $\sigma^{i}_{x,y}\to - \sigma^{i}_{x,y}$ for $i=1,3,...$. This ambiguity in the phase choice does not affect any of the results presented in the main text, since the observables we calculate depend on the absolute values of $a_k$. However, we note that for bosonic systems the use of $H_{|\Psi_0|}$ instead of $H$ can facilitate calculations of such quantities as correlations functions etc., since  $|\Psi_0|$ is symmetric with respect to two-body exchanges. Other choices of the phase in Eq. (\ref{wf-interacting}), e.g., $\Psi_0 \to e^{i \phi_k} \Psi_0, \phi_k \in \mathbb{R}$,  lead to other spin-chain Hamiltonians that are unitarily equivalent.

For conditional spin transfer in a four-particle spin chain, we require that 
the exchange coefficients satisfy $J_{1,3} / J_2 \ll 1$, which can be realized 
in a symmetric triple-well potential $V(x)$ given by
\begin{equation}
\label{seq:1Dpot}
V(x) = -V_0 \left[e^{-a (x-x_0)^2}+ e^{-a (x+x_0 )^2}\right]  - U e^{-b x^2}    ,
\end{equation}
where $V_0$ and $U$ are in units of $\varepsilon = \frac{1}{mL^2}$.
We set the centers of the spatial Gaussians with
$x_0 = \frac{7 L}{16}$ and choose the constants 
$a = \frac{384}{L^2}$ and $b = \frac{64}{5L^2}$.

The shape of the potential for values $V_0 = 500$ and $U = 200$ is shown in Fig.~2
of the main text. The potential wells next to the edges are very deep 
compared to the broad well in the middle. The single-particle wavefunctions
are also schematically shown there. We observe that the two lowest-energy 
eigenstates are located almost fully in the deep wells next to the boundaries 
of the potential. In the inset, we show the ratio of exchange coefficients 
$J_2 / J_1$ as a function of the parameter $U$ in the potential in Eq.~(\ref{seq:1Dpot}). 
This ratio can reach rather large values, for example $J_2 / J_1 > 10$ for 
$U = 150-550$. This potential can therefore be used to attain the necessary 
parameters of the $XXZ$ model Hamiltonian to realize conditional spin transfer.

\end{document}